\documentclass[pra,twocolumn,superscriptaddress]{revtex4}
\usepackage{blindtext}
\usepackage{amsfonts}
\usepackage{amsmath,amssymb}
\usepackage{graphicx}
\usepackage{subfigure}
\usepackage{bm}
\usepackage{epstopdf}
\usepackage{url}
\usepackage{hyperref}
\hypersetup{
     colorlinks = true,
     linkcolor = blue,
     anchorcolor = blue,
     citecolor = blue,
     filecolor = blue,
     urlcolor = blue
}

\begin{document}

\title{Localization of Pairs in One-Dimensional Quasicrystals with Power-Law Hopping}
\author{G. A. Dom\'{\i}nguez-Castro and R. Paredes} 
\affiliation{Instituto de F\'{\i}sica, Universidad
Nacional Aut\'onoma de M\'exico, Apartado Postal 20-364, M\'exico D.F. 01000, Mexico.}  
\email{rosario@fisica.unam.mx}

\begin{abstract}
Pair localization in one-dimensional quasicrystals with nearest-neighbor hopping is independent of whether short-range interactions are repulsive or attractive. We numerically demonstrate that this symmetry is broken when the hopping follows a power law $1/r^{\alpha}$. In particular, for repulsively bound states, we find that the critical quasiperiodicity that signals the transition to localization is always bounded by the standard Aubry-Andr\'e critical point, whereas attractively bound dimers get localized at larger quasiperiodic modulations when the range of the hopping increases. Extensive numerical calculations establish the contrasting nature of the pair energy gap for repulsive and attractive interactions, as well as the behavior of the algebraic localization of the pairs as a function of quasiperiodicity, interaction strength, and power-law hops. The results here discussed are of direct relevance to the study of the quantum dynamics of systems with power-law couplings. 
\end{abstract}

\maketitle

\section{Introduction}
\label{SecI}
Quasicrystals are intriguing structures that are characterized by having long-range order without spatial periodicity. Such exotic states of matter constitute intermediate cases between disordered and periodic systems. Due to their particular spatial arrangement, the localization of individual particles in one-dimensional quasicrystals has been explored both theoretically \cite{Aubry1, Macia1, Sarma1, Modugno1, Tanatar1, Logan1, Sarma2} and experimentally \cite{Roati1, Bloch1, Lahini1}. In particular, considerable attention has been given to the celebrated Aubry-Andr\'e (AA) model \cite{Aubry1, Harper1, Azbel1}. In this model, quasiperiodicity emerges by superimposing two lattices with incommensurate periods \cite{Gustavo1}, and particles hop through nearest-neighbor sites only. To enrich the localization problem in the AA model, one can replace the nearest-neighbor tunneling with a hopping whose amplitude follows a power law. This modification is of particular interest since power-law interactions arise in many important systems \cite{Macri1}. For instance, polar molecules \cite{DeMarco1, Moses1, Yan1}, Rydberg atoms \cite{Browaeys1, Browaeys2}, trapped ions \cite{Blatt1, Blatt2, Monroe1}, and atoms in photonic crystal waveguides \cite{Kimble1}, among others. Intriguing results, such as multifractal states \cite{Santos1,Roy1} and algebraic localization \cite{Santos2}, arise as a consequence of the interplay between quasiperiodicity and power-law hops.

One of the fundamental questions in localization theory, which has sparked intense debate \cite{Scarola1, Panda1, Vidmar1, Sirker1, Sirker2}, is the fate of the Anderson transition in the presence of interactions at finite particle density. This subject, usually called the many-body localization problem \cite{Huse1}, faces significant computational and experimental challenges. From the numerical side, exact diagonalization methods are restricted to small size systems due to the exponential growth of the Hilbert space, tensor network algorithms \cite{Vidal1, Ulrich1, Verstraete1} allow one to simulate the dynamics of larger systems, but up to times limited by the amount of entanglement in the many-body system. On the other hand, experiments are restringed to several hundred tunneling times due to the coupling with the external environment \cite{Bloch2, Bloch3, Bloch4, Bloch5}. This limitation makes it difficult to extrapolate the results to the infinite time limit, where a slow decay regime can be straightforwardly distinguished from the peculiar frozen dynamics of MBL. 

Because of the complexity of the many-body localization problem, it is reasonable to focus our attention on the localization properties of few-body systems. Although it is true that the collective behavior of matter demands the participation of a large conglomerate of entities, the physics of two or few interacting particles can contain the essence from which one can understand the properties of a many-body system. In fact, despite its apparent simplicity, the pair localization problem already exhibits rich physics. For instance, the enhancement of the pair localization length \cite{Shepelyansky1, Oppen1, Flach1, Flach2}, the interaction effect on the dimer localization  \cite{Orso1, Flach3, Frahm1, Shepelyansky2, Kim1, Mujal1, Liu1, Rai1}, the presence or absence of mobility edges  \cite{Orso2, Orso3}, the fractal character of the two-body spectrum \cite{Andreanov1}, and exotic dynamical regimes \cite{Gustavo2}, among others. Furthermore, due to the high precision and tunability achieved on several quantum simulation platforms, the observation of few-body phenomena is within the reach of current experiments \cite{Jochim1, Jochim2, Jochim3}.

In this manuscript, we study a fundamental two-body model that incorporates the essential ingredients discussed above: short-range interactions, quasiperiodic potential, and power-law hopping. To the best of our knowledge, there is no study that discusses the interplay of the three former elements on the pair localization transition. We numerically demonstrate that, in stark contrast with quasicrystals with nearest-neighbors hops \cite{Shepelyansky1, Flach1, Orso1, Frahm1, Rai1}, the mirror symmetry of the critical quasiperiodicity, where localization occurs, breaks when the hopping range is increased. That is, its value depends on whether the interactions are attractive or repulsive. Our calculations show that the critical quasiperiodic modulation for repulsively bound states is always bounded by the usual Aubry-Andr\'e transition point \cite{Aubry1, Gustavo1}, whereas attractively dimers localize at larger quasiperiodic modulations when the range of the hopping increases. Furthermore, through extensive numerical calculations, we study the pair energy gap and the algebraic decay of both, repulsively and attractively bound dimers. The results here discussed go beyond previous findings \cite{Shepelyansky2, Flach1, Orso1}, in the sense that they explore the consequences of the range of the hopping on the two-body localization transition. Moreover, are of main relevance for current studies on the quantum dynamics of bound states in optical lattices \cite{Gustavo2, Macri2, Santos3, Santos4}.

The manuscript is organized as follows. In Sec. \ref{SecII} we introduce the model considered and develop the Green's function formalism used to address the two-body problem. Subsequently, in Sec. \ref{SecIII} we numerically demonstrate the extended-localized transition and calculate localization properties for both, attractively and repulsively dimers. Finally, in Sec. \ref{SecIV}, we summarize and conclude the manuscript.

\section{Green's function formalism}
\label{SecII}

We consider a pair of interacting particles moving in a one-dimensional quasicrystal with power-law hopping. The Hamiltonian of the two-body system can be written as $\hat{H} = \hat{H}_{0} + \hat{U}$, with $\hat{H}_{0}$ the noninteracting component and $\hat{U}$ the interaction operator. The ideal part of $\hat{H}$ can be decomposed as $\hat{H}_{0} = \hat{H}_{\text{GAA}}\otimes\mathbb{I}_{1} + \mathbb{I}_{1}\otimes\hat{H}_{\text{GAA}}$, being $\mathbb{I}_{1}$ the one-body identity operator and $\hat{H}_{\text{GAA}}$ the single-particle generalized Aubry-Andr\'e Hamiltonian:
\begin{equation}
\begin{split}
\hat{H}_{\text{GAA}} =  -J\sum_{i,j\neq i} & \frac{1}{|i-j|^{\alpha}}|i\rangle\langle j|\\ &  + \Delta\sum_{j}\cos(2\pi\beta j + \phi)|j\rangle\langle j|,
\end{split}
\label{Eq1}
\end{equation}
where $|j\rangle$ stands for the state in which the particle is localized at the site $j$, and $J/|i-j|^{\alpha}$ is the hopping rate between the sites $i$ and $j$. The quasiperiodic modulation is characterized by its strength $\Delta$, the incommensurate parameter $\beta = (\sqrt{5}-1)/2$, and a random phase $\phi\in[0,2\pi)$. For $\alpha\gg 1$, the GAA model approaches to the celebrated AA model \cite{Aubry1, Harper1, Azbel1}. As it is well-known, all the eigenstates of the Aubry-Andr\'e Hamiltonian are extended for $\Delta/J<2$, all localized for $\Delta/J>2$, and all multifractal at the transition point $\Delta/J=2$ \cite{Wilkinson1}. In contrast, the GAA model displays a plethora of mobility edges that split extended and localized single-particle states for $\alpha>1$ \cite{Santos1}, whereas for long-range hops $\alpha<1$, the single-particle states are extended or multifractal \cite{Santos1, Roy1}. The operator $\hat{U}$ couples the two particles by an onsite interaction of strength $U$:
\begin{equation}
\hat U= U\sum_j |j,j\rangle \langle j, j |,
\label{Eq2}
\end{equation}
being $|j ,j \rangle = |j\rangle\otimes|j\rangle$ the two-body state in which both particles are in the lattice site $j$. It is important to mention that the onsite interaction in Eq. (\ref{Eq2}) plays a role for spatially symmetric wave functions, where the particles can be found on the same site with nonzero probability. Thus, our results are relevant when the particles are bosons or fermions with opposite spins in the singlet state. The Schr\"odinger equation for the two-particle state $|\Psi \rangle$ can be written as $(E-\hat{H}_0) |\Psi\rangle= \hat{U}|\Psi \rangle$ with $E$ the energy. This equation can be numerically solved with the aid of the noninteracting two-body Green's function operator $\hat{\mathcal{G}}_{0}(E)= (E-\hat {H}_0)^{-1}$ which, can be formally written in terms of the eigenstates $|\varphi_{l}\rangle$ and eigenenergies $\varepsilon_{l}$ of the GAA Hamiltonian:
\begin{equation}
\hat{\mathcal{G}}_{0}(E) = \sum_{n,m}\frac{1}{E-\varepsilon_{n}-\varepsilon_{m}} |\varphi_{n},\varphi_{m}\rangle\langle\varphi_{n},\varphi_{m}|.
\label{Eq3}
\end{equation}
By applying $\hat{\mathcal{G}}_{0}$ to both sides of the Schr\"odinger equation $(E-\hat{H}_0) |\Psi\rangle= \hat{U}|\Psi \rangle$, one can find $|\Psi\rangle=\hat{\mathcal{G}}_{0}\hat{U}|\Psi\rangle$. Projecting this last expression over the state $|j, j' \rangle$, we obtain an equation for the amplitudes $\Psi(j,j')=\langle j,j'|\Psi\rangle$ of the two-particle wave function:
\begin{equation}
\begin{split}
\Psi(j,j') &= \langle j, j'|\hat{\mathcal{G}}_{0}\hat{U}|\Psi\rangle  \\ &=  U\sum_{i}\langle j, j'|\hat{\mathcal{G}}_{0}(E)|i, i\rangle\Psi(i,i),
\end{split}
\label{Eq4}
\end{equation}
where in the last equality we use the fact that the interaction operator $\hat{U}$ is diagonal in the space representation. The Eq. (\ref{Eq4}) shows that $\Psi(j,j')$ is entirely determined by its diagonal components $\Psi(j,j)$ which, for simplicity, we shall denote by $\psi(j)=\Psi(j,j)$. Setting $j=j'$ in Eq. (\ref{Eq4}) yields the desired eigenvalue problem:
\begin{equation}
\frac{1}{U}\psi(j) = \sum_{i} G_{0}(j,i;E)\psi(i),
\label{Eq5}
\end{equation} 
being $G_{0}(j,i;E)$ the matrix elements of the noninteracting two-body Green's function operator:
\begin{equation}
\begin{split}
G_{0}(j,i;E) =& \langle j, j|\hat{\mathcal{G}}_{0}(E)|i, i\rangle\\
=& \sum_{n,m}\frac{\varphi_{n}(j)\varphi_{m}(j)\varphi_{n}^{*}(i)\varphi_{m}^{*}(i)}{E-\varepsilon_{n}-\varepsilon_{m}}.
\end{split}
\label{Eq6}
\end{equation}
The computational complexity of the above equation is O$(L^{4})$ and, in contrast with tight-binding lattices, it cannot be reduced to O$(L^{3})$ since $\hat{H}_{GAA}$ does not have a tridiagonal structure. For this reason, we restrict our calculations to a moderate lattice size of $L=377$. Since the dimer motion is confined to one dimension, the bound state exists for arbitrarily small interactions. Furthermore, the eigenvalue problem in Eq. (\ref{Eq5}) admits solutions for negative and positive interaction stregths. The former states are called attractively bound pairs and the latter repulsively bound pairs. In contrast to dimers bounded by an attractive interaction, a repulsively bound pair is not the ground state of the two-body system. However, due to energy constraints, the repulsively bound dimer is unable to decay by converting the interaction energy into kinetic energy and is therefore dynamically stable. In this manuscript, we concentrate in the maximal an minimal energy states of Eq. (\ref{Eq5}). In the absence of quasiperiodic modulation, these states correspond to a pair of particles with zero centre-of-mass momentum. To avoid any inconvenience with the thermodynamic limit of the pair energy, throughout the manuscript we consider $\alpha>1$ only.

\section{Results}
\label{SecIII}

\subsection{Extended-localized transition}

\begin{figure}[h]
\includegraphics[width=3.2in]{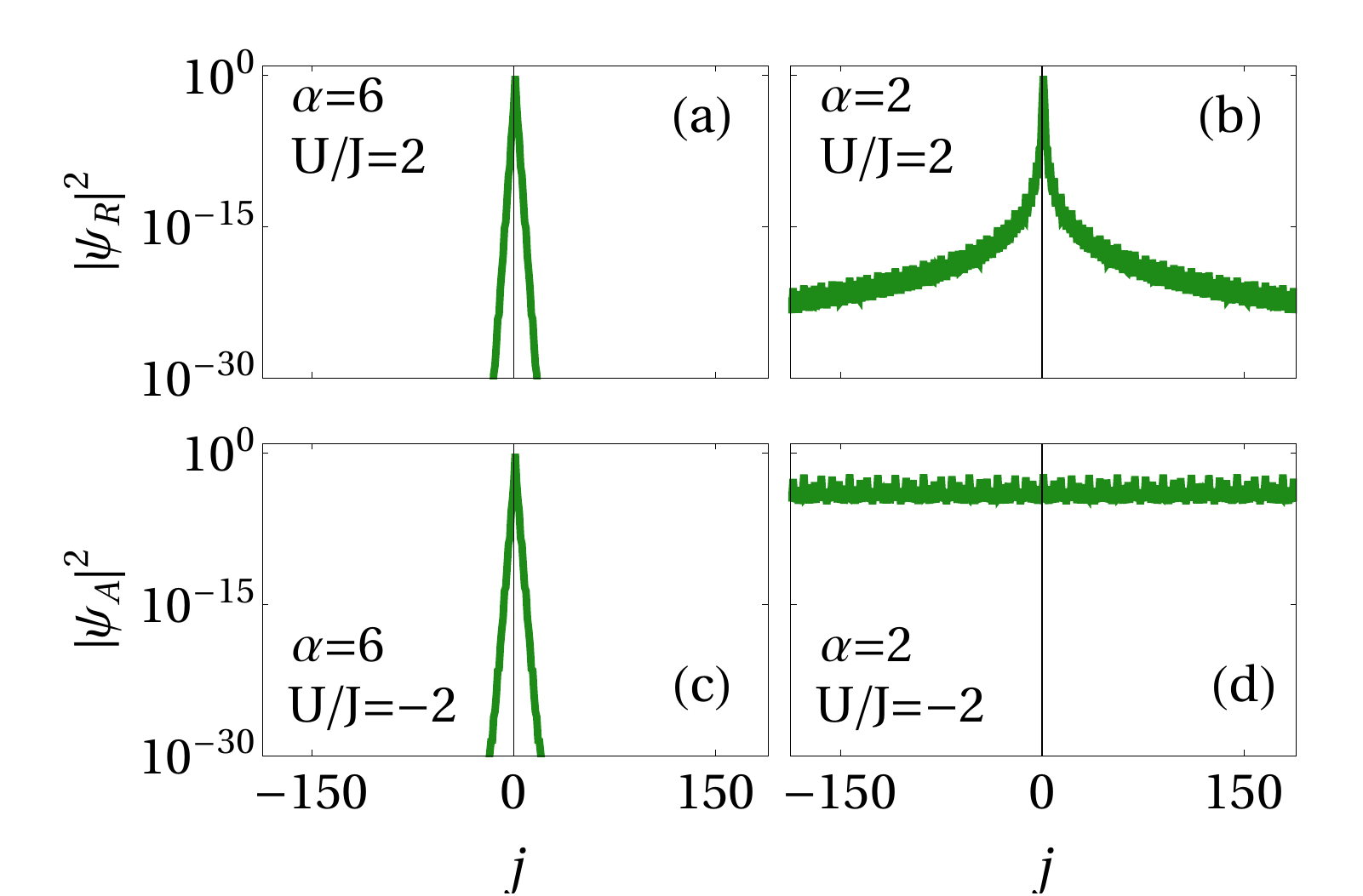}
\caption{Diagonal elements of the two-body profile $\psi(j)=\Psi(j,j)$ as a function of the lattice index $j$, $\psi_{A}$ and $\psi_{R}$ correspond to attractively and repulsively bound pairs, respectively. The strength of the quasiperiodic modulation for all panels is $\Delta/J=2$.}
\label{Fig1}
\end{figure}

In one-dimensional quasicrystals with nearest-neighbor tunneling, the wave function of an attractively bound state with energy $E$ describes also a repulsively bound state with energy $-E$, provided the phase $\phi$, belonging to the AA potential, is shifted by $\pi$. In other words, while the the attractively bound state localizes at the minimum of the quasiperiodic modulation, the repulsively bound state gets localized at the maximum. The fact that both kinds of pairs are represented by the same spatial profile implies that the two-body extended-localized transition does not depend on the interaction sign \cite{Orso1}. That is, the critical quasiperiodicity $\Delta_{c}$ at which the transition takes place is an even function of the interaction strength $\Delta_{c}(-U) = \Delta_{c}(U)$. This is no longer true for one-dimensional quasicrystals with power-law hops, as illustrated in Fig. \ref{Fig1} (see the log scale in the vertical axis), the diagonal elements of the two-body wave functions can show contrasting spatial behaviors. In particular, while the attractively bound pair is extended for $\alpha=2$ and $(U/J,\Delta/J)=(-2,2)$, the associated repulsively bound state for $U/J=2$ is localized. For $\alpha=6$, the pair states are nearly identical, as expected from quasicrystals with short-range hops. The localized wave functions plotted in Fig. \ref{Fig1} correspond to phases $\phi$ suitably chosen so that the localization center coincides with the middle of the lattice.   

\begin{figure}[h]
\includegraphics[width=3.3in]{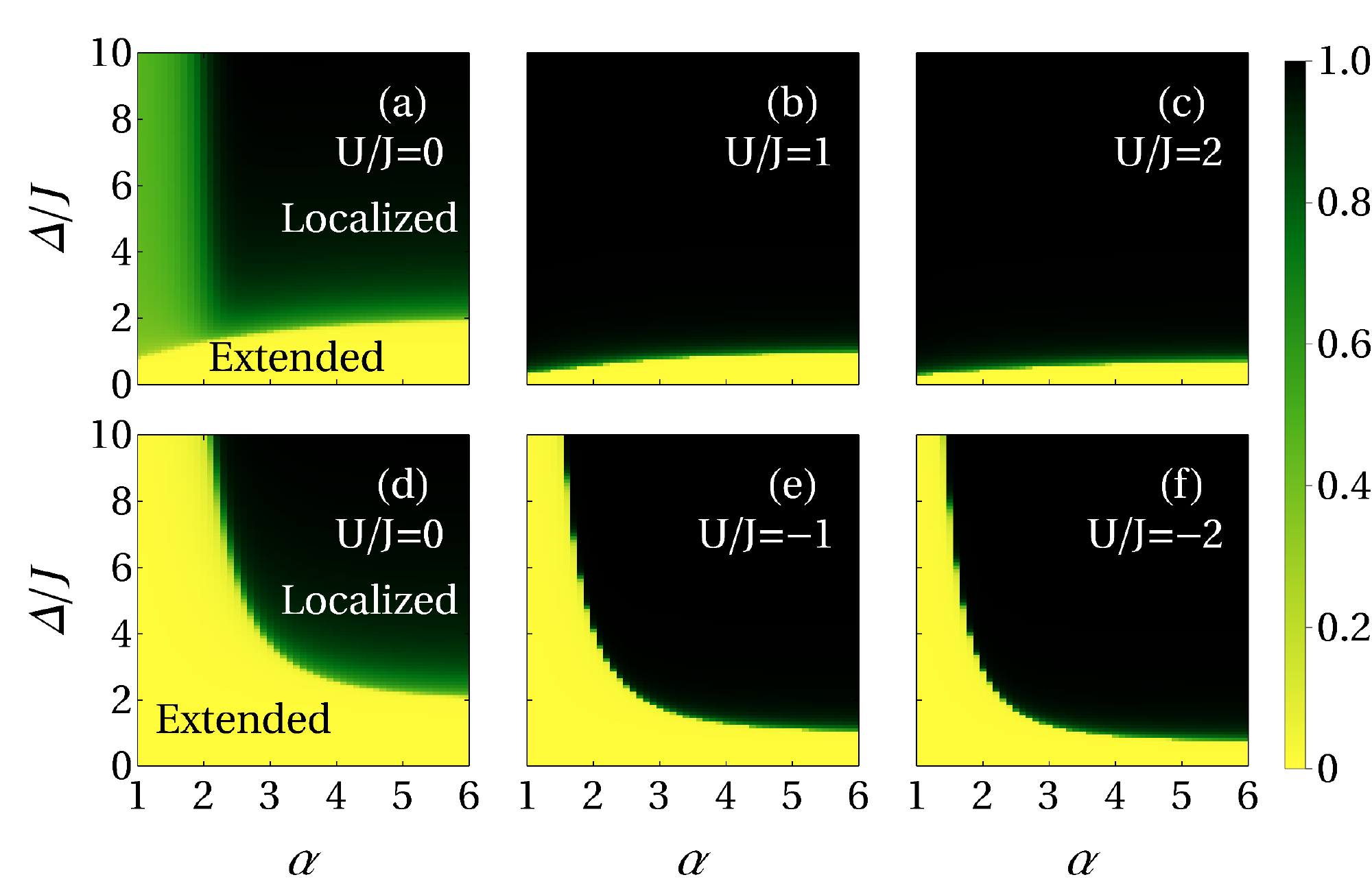}
\caption{Inverse participation ratio as a function of the power hop $\alpha$ and the quasiperiodicity $\Delta/J$ for several interaction strengths. The noninteracting cases in panels (a) and (d) are associated with the maximal and minimal scattering states, respectively. All the calculations for finite quasiperiodicity were obtained from the average of $30$ random uniformly distributed phases $\phi\in[0,2\pi)$.}
\label{Fig2}
\end{figure}

A customary parameter that is used as a measure of localization is the inverse participation ratio, given a normalized wave function $\psi$ its IPR is defined as IPR$_{\psi} = \sum_{i=1}^{L}|\psi(i)|^{4}$. For extended states, the IPR vanishes in the thermodynamic limit as $\propto L^{-1}$, whereas for spatial localized profiles is always finite. In Fig. \ref{Fig2}, we plot in a density color scheme the inverse participation ratio of attractively and repulsively bound dimers as a function of $\alpha$ and $\Delta/J$ for several interaction strengths. The noninteracting cases shown in Fig. \ref{Fig2}(a) and Fig. \ref{Fig2}(d)  correspond to the maximal and minimal energy states of the scattering band, respectively. As one can notice, the interaction between particles favors the localization of both kinds of pairs. However, the IPR shows distinct features for attractive and repulsive interactions. For instance, as long as $\alpha\lesssim 2$ attractively bound dimers are extended for large quasiperiodic modulations $\Delta/J\approx 8$. In contrast, all repulsively bound states are localized for these parameters. It is interesting to note that for attractively bound pairs, the extended region, where the IPR is null, enlarges as $\alpha$ decreases. In contrast, the extended region for repulsively bound dimers enlarges as $\alpha$ increases.

To determine the critical quasiperiodicity $\Delta_{c}/J$ at which the localization transition of the pairs takes place, we employ the inflection point of the inverse participation ratio. That is, for fixed values of $U$ and $\alpha$, we calculate the IPR of the dimer state $\psi$ as a function of $\Delta$ then, we find the point where the second derivative of the obtained curve IPR$(\Delta)$ vanishes. The inflection point technique has been successfully used in several previous works \cite{Orso1, Liu1}.
\begin{figure}[h]
\includegraphics[width=3.0in]{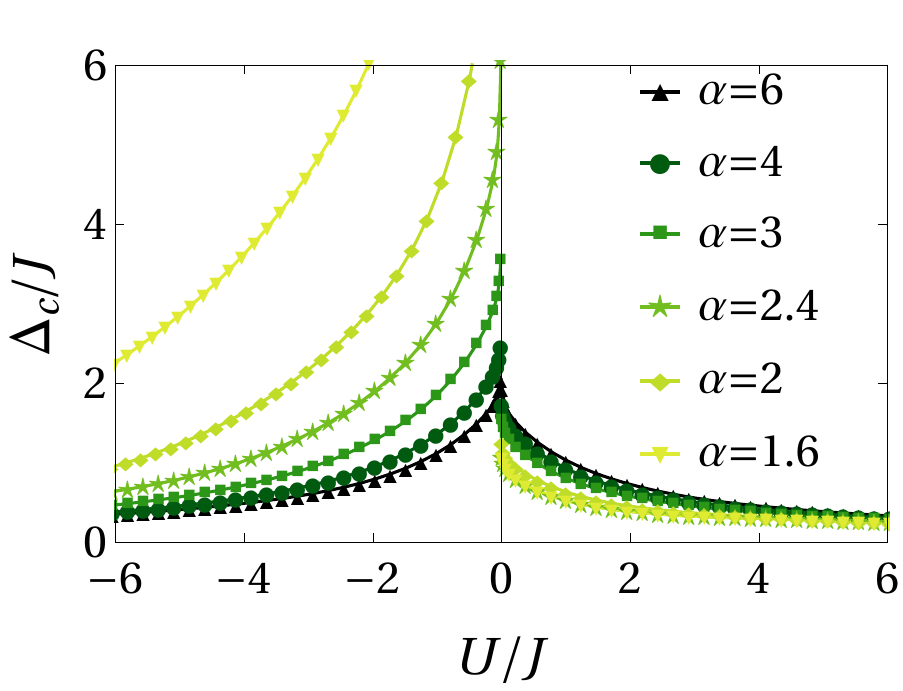}
\caption{Critical quasiperiodicity of the two-body extended-localized transition as a function of the interaction strength for several power-law hops. All the calculations were obtained from the average of $30$ random uniformly distributed phases $\phi\in[0,2\pi)$.}
\label{Fig3}
\end{figure}

\begin{figure}[h]
\includegraphics[width=2.5in]{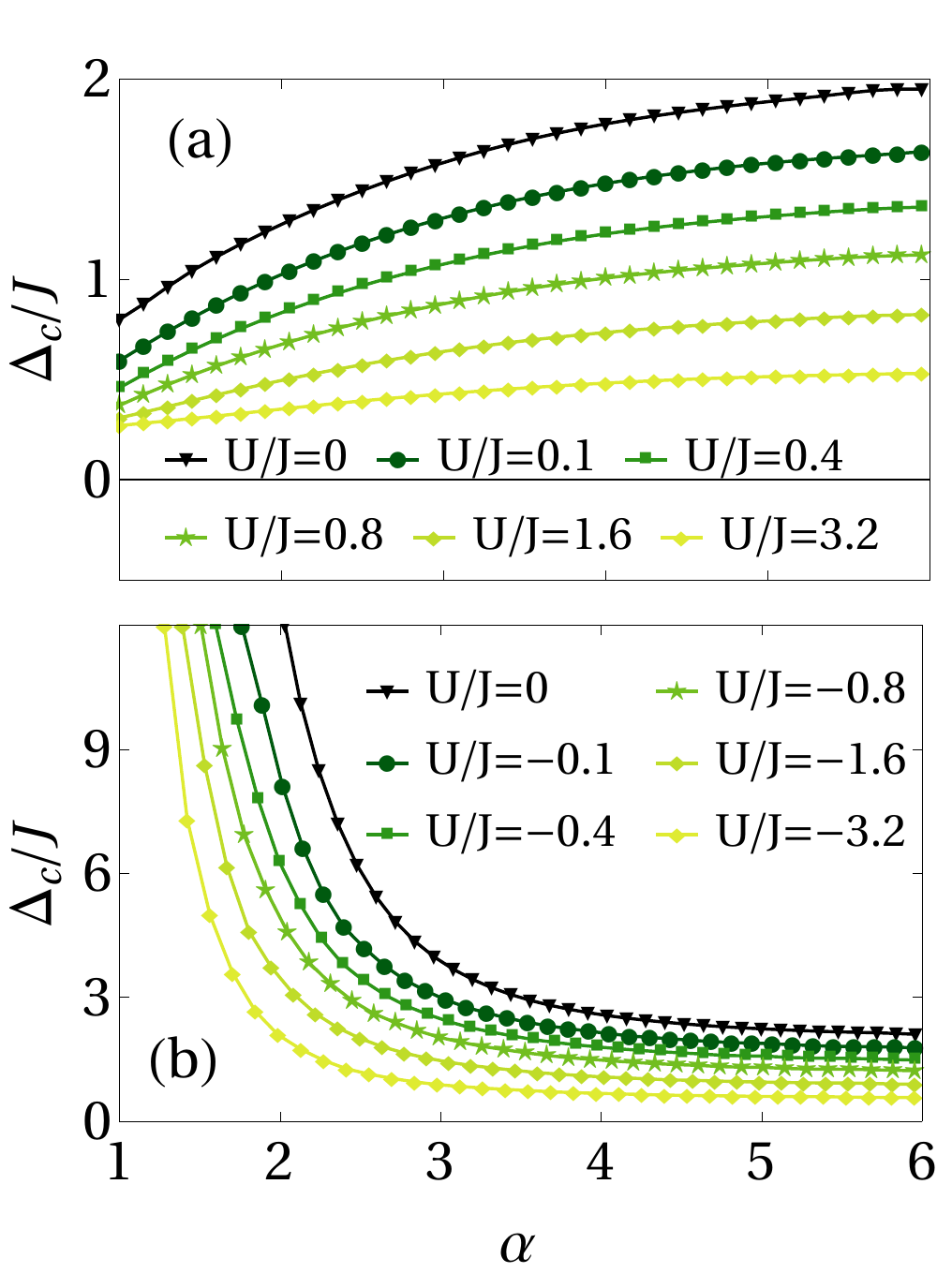}
\caption{Critical quasiperiodicity $\Delta_{c}/J$ as a function of the hopping range $\alpha$ for several interaction strengths $U/J$. Panels (a) and (b) correspond to repulsively and attractively bound dimers, respectively. All the calculations were obtained from the average of $30$ random uniformly distributed phases $\phi\in[0,2\pi)$.}
\label{Fig4}
\end{figure}

In Fig. \ref{Fig3} we illustrate the critical quasiperiodicity $\Delta_{c}/J$ of the dimer localization transition as a function of the interaction strength $U/J$ for several values of the power hop $\alpha$. For $\alpha=6$, the critical quasiperiodic modulation is approximately an even function of the interaction strength, in agreement with quasicrystals with short-range hops \cite{Orso1}. As $\alpha$ decreases, the mirror symmetry of $\Delta_{c}/J$ with respect to $U/J$ is completely broken, namely $\Delta_{c}(-U) \neq \Delta_{c}(U)$. In particular, attractively bound dimers get localized at larger quasiperiodic modulations than that of the repulsively bound states. Since in the strong coupling regime where the particles are tightly bound, $|U|\gg J, \Delta$, the effective mobility of the dimer follows a power-law $1/r^{\sigma}$, with $\sigma=2\alpha$, (see Appendix \ref{AppendixA}), namely is of shorter range than the hops of individual particles, the mirror symmetry of $\Delta_{c}/J$ is slowly restored. To analyze the behavior of the two-body localization transition with respect to the range of the hopping, in Fig. \ref{Fig4}, we illustrate the behavior of $\Delta_{c}/J$ as a function of $\alpha$ for several values of the interaction strength. Notice that the critical quasiperiodicity for repulsively bound pairs is always bounded by the Aubry-Andr\'e transition point $\Delta_{c}/J=2$. In contrast, for attractive interactions, $\Delta_{c}/J$ exceeds the AA bound when the range of the hopping increases.

\subsection{The pair energy gap}
\begin{figure}[h]
\includegraphics[width=3.64in]{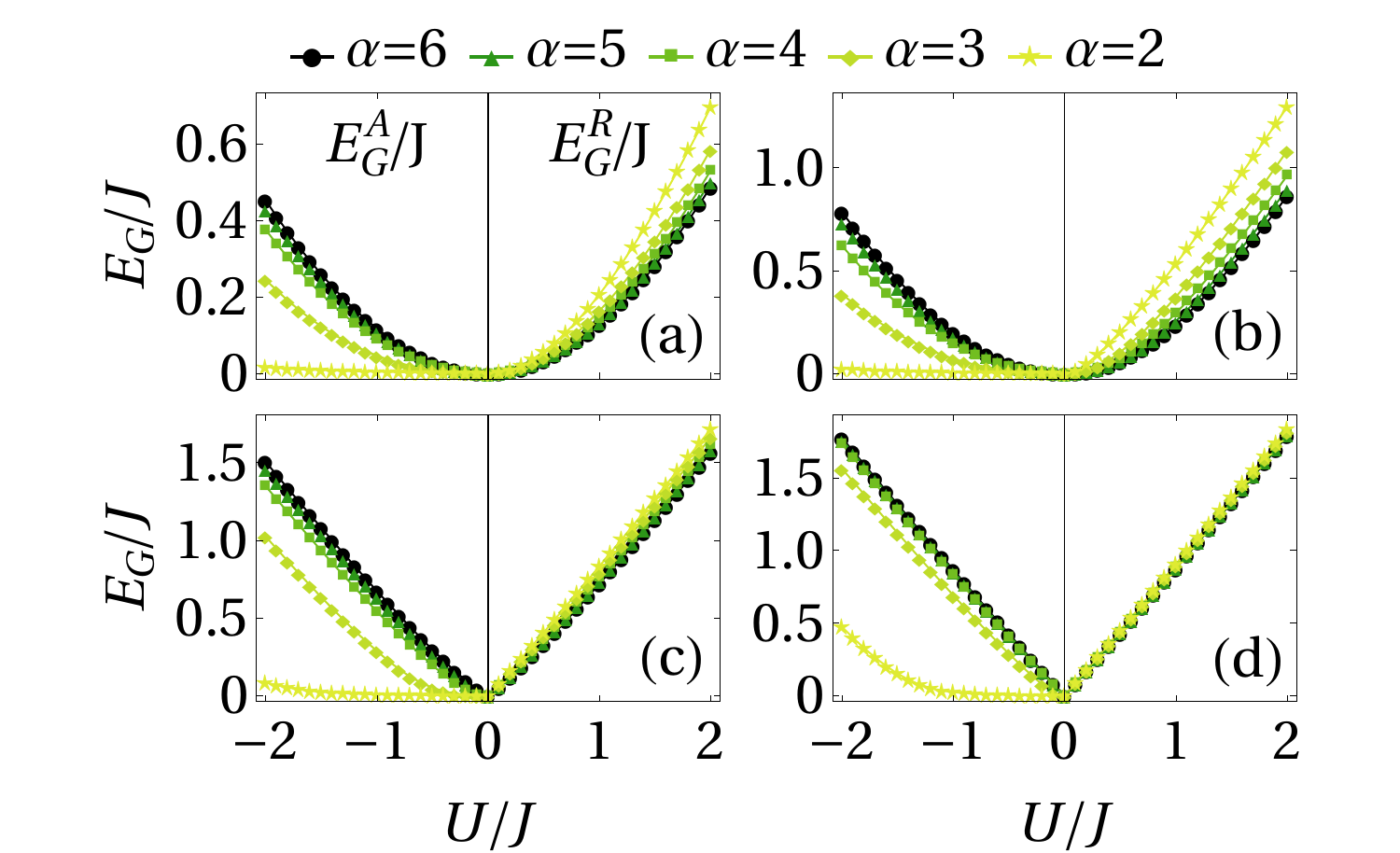}
\caption{The pair energy gap as a function of the interaction strength $U$ for several values of the power-law hopping $\alpha$.
$E_{G}^{A}$ and $E_{G}^{R}$  correspond to the attractive and repulsive branches, respectively. Panels (a), (b), (c) and (d) are associated with a quasiperiodic strength of $\Delta/J=0, 1, 2$ and $3$, respectively.}
\label{Fig5}
\end{figure}

As it is well known, the total energy of repulsively and attractively bound pairs lies above and below the two-body scattering energies, respectively. The energy gap $E_{G}$ between a bound state and the closest scattering state is a measure of the required energy to dissociate the pair. Furthermore, the pair energy gap can be measured experimentally in optical lattices using rf spectroscopy \cite{Martin1, Tilman1}. Mathematically, $E_{G}$ is defined as follows:
\begin{equation}
\begin{split}
E_{G}^{R} &= E-2\varepsilon_{L}\\
E_{G}^{A} &= 2\varepsilon_{1}-E,
\end{split}
\label{Eq7}
\end{equation}
where the superscripts R and A are associated with repulsively and attractively bound dimers, respectively, $\varepsilon_{1}$ is the lowest energy and $\varepsilon_{L}$ is the highest energy of the single-particle spectrum. In Figs. \ref{Fig5}(a)-\ref{Fig5}(d), we show the pair energy gap $E_{G}$ as a function of the interaction strength for $\Delta/J=0, 1, 2$ and $3$, respectively. The values of the hopping power $\alpha$ are indicated in different colors. As illustrated in Fig. \ref{Fig5}(a), $E_{G}^{R}$ correspond to positive $U$, while $E_{G}^{A}$ is associated with negative interactions. For the periodic lattice $\Delta/J=0$, one can see a very nearly mirror image between $E_{G}^{A}$ and $E_{G}^{R}$ when $\alpha=6$. However, as $\alpha$ decreases, the asymmetric behavior of the pair energy gap is clearly seen. In particular, repulsively bound states exhibit a larger $E_{G}$ than the associated with attractively bound pairs. Because localized profiles increase the interaction energy, one can recognize from Figs. \ref{Fig5}(b)-\ref{Fig5}(d) that the pair energy gap for both kinds of dimers increases significantly when the quasiperiodic potential localizes the two-particle wave function. Furthermore, due to a strong localization, the mobility of the pair ceases to play a relevant role, and therefore the curves associated with different values of $\alpha$ gradually collapse to a straight line.

\subsection{Algebraic localization}
\begin{figure}[h]
\includegraphics[width=3.0in]{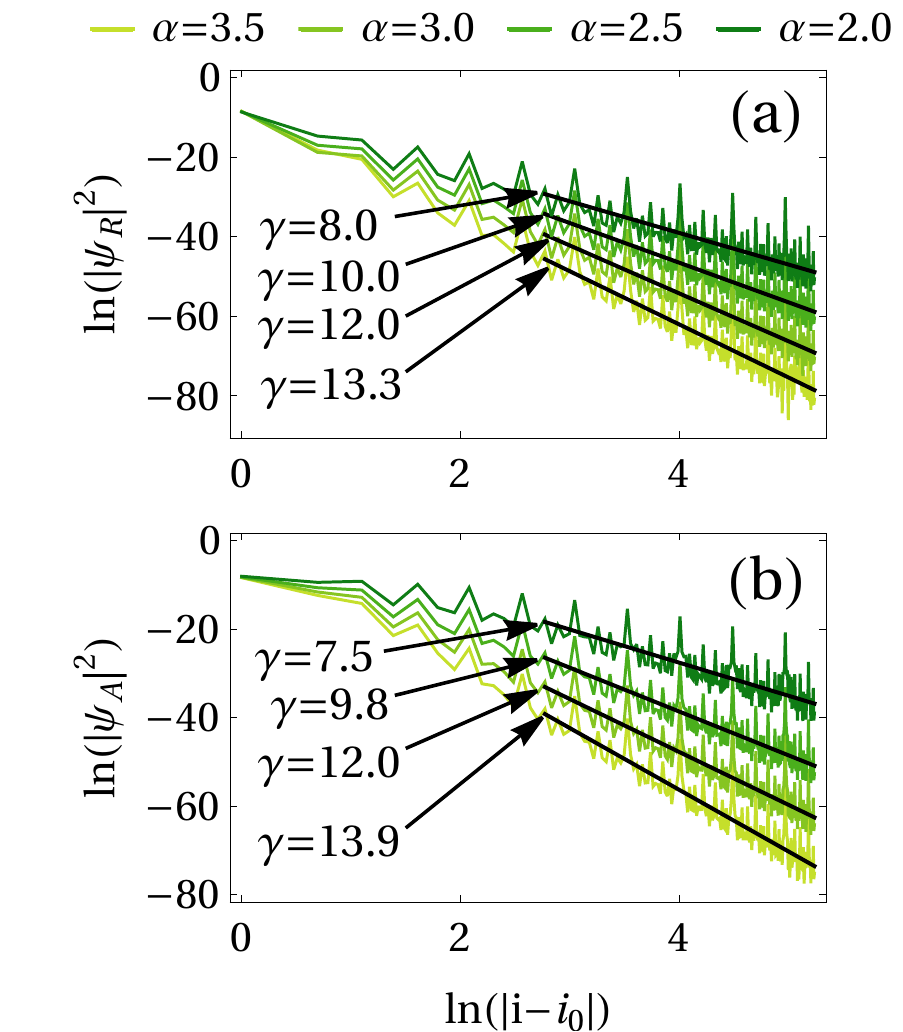}
\caption{Logarithm of $|\psi|^{2}$ vs logarithm of the distance $i$ from the localization center $i_{0}$.
The arrows indicate the decay power $|\psi|^{2} \sim |i-i_{0}|^{-\gamma}$ of each wave function, $\psi_{A}$ and $\psi_{R}$ correspond to attractively and repulsively bound dimers, respectively. The parameters are $(U/J, \Delta/J) = (3,4)$ for panel (a), whereas panel (b) is associated with $(U/J, \Delta/J) = (-3,4)$.}
\label{Fig6}
\end{figure}

As it is well-known, localization in quasicrystals with nearest-neighbor hopping is characterized by exponential tails $|\psi(i)|^{2} \sim e^{-|i-i_{0}|/\xi}$, being $\xi$ the localization length. In contrast, power-law tunneling yields algebraic decay $|\psi(i)|^{2} \sim |i-i_{0}|^{-\gamma}$, where $\gamma$ is the decay power and $i_{0}$ the localization center, which is placed at the maximum value of $|\psi|^{2}$. Recently, it has been found that algebraic single-particle states can be either conducting or insulating in the thermodynamic limit \cite{Saha1}. As illustrated in Fig. \ref{Fig6}, the spatial distribution of attractively and repulsively dimers is well fitted by the algebraic dependence $|i-i_{0}|^{-\gamma}$, the arrows in each panel indicate the corresponding value of $\gamma$. As one can notice, the decay power for both kinds of pairs increases as the tunneling range of the particles decreases. That is, $|\psi|^{2}$ falls off more abruptly in space when $\alpha$ increases. It is important to comment that for $\alpha \gtrsim 4$, we found that the algebraic ansatz fits the profile poorly because $|\psi|^{2}$ gradually recovers its exponential tail. 

\begin{figure}[h]
\includegraphics[width=3.0in]{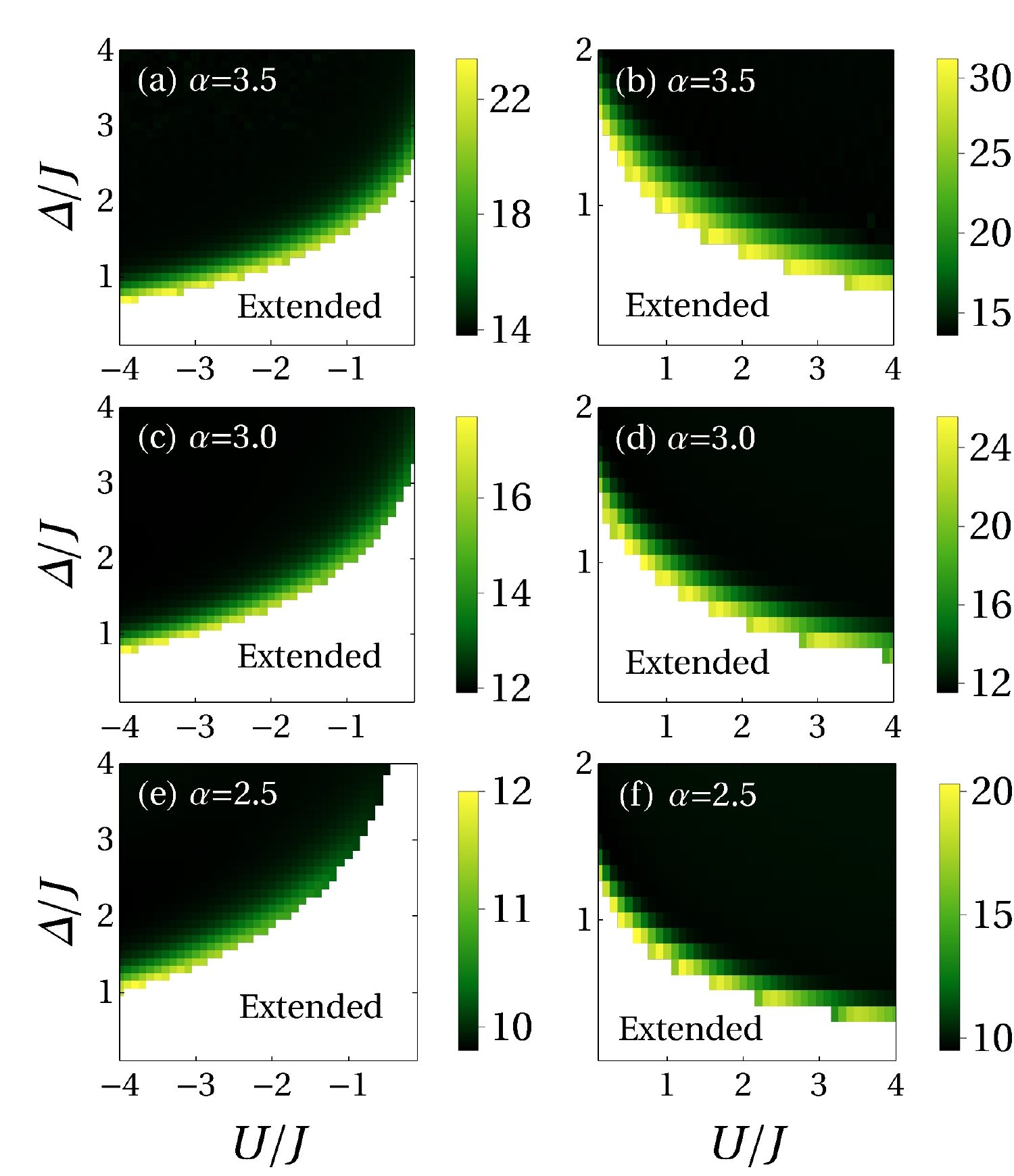}
\caption{Decay power $\gamma$ as a function of interaction $U/J$ and quasiperiodicity $\Delta/J$ for two different power-law hops. Panels (a) and (b) are associated with repulsively bound states whereas panels (c) and (d) correspond to attractively dimers. All the calculations were obtained from the average of $30$ random uniformly distributed phases $\phi\in[0,2\pi)$.}
\label{Fig7}
\end{figure}

Figures \ref{Fig7}(a)-\ref{Fig7}(f) show the value of the decay power $\gamma$ as a function of $\Delta/J$ and $U/J$ for repulsively and attractively bound pairs with three different values of $\alpha$. As one can notice, deep in the localized regime $\gamma$ is approximately constant for both kinds of dimers. However, near the transition point, the decay power varies strongly. In particular, we find that close to $\Delta_{c}$, the wave function of repulsively bound dimers decays faster in space than that associated with attractively bound states.

\section{Conclusion}
\label{SecIV}

We have investigated the localization properties of two interacting particles moving in a one-dimensional quasicrystal with an adjustable tunneling range. In the proposed model, two identical bosons or fermions with opposite spins are coupled via a short-range interaction and tunnel not just through nearest-neighbor sites, but across the whole lattice with hopping couplings that follows a power-law function $1/r^{\alpha}$. By using Green's function techniques and numerical exact diagonalization, we have found that, in stark contrast with pair localization in quasicrystals with nearest-neighbor hops \cite{Orso1}, the extended-localized transition of the dimer strongly depends on whether the interaction is repulsive or attractive. That is, the mirror symmetry of the critical quasiperiodicity at which the transition takes place is broken. In particular, we showed that the critical quasiperiodic modulation for repulsively bound states is always bounded by the usual Aubry-Andr\'e transition point, whereas attractively pairs localize at larger quasiperiodic strengths when the range of the hopping increases. Furthermore, we numerically demonstrated that as the hopping range is decreased, the mirror symmetry of the critical quasiperiodic modulation is restored, in agreement with previous literature on quasicrystals with nearest-neighbor hops. An extensive set of numerical calculations, allowed us to determine the effects of interactions, quasiperiodicity, and hopping range on both, the pair energy gap as well as the algebraic localization of the two-body system.

We expect that our analysis will trigger further theoretical work in determining both, the fate and effects of dimer formation in the transport properties of many-body systems with power-law couplings. The model proposed in our manuscript is of current relevance for several quantum simulation platforms where power-law interactions emerge. For instance, trapped ions, Rydberg atoms, polar molecules, and atoms in photonic crystal waveguides among other systems.

\acknowledgments{This work was partially funded by Grant No. IN108620 from DGAPA (UNAM). G.A.D.-C acknowledges CONACYT scholarship.}

\appendix
\section{Effective Hamiltonian}
\label{AppendixA}

In this appendix we introduced an effective Hamiltonian that describes the two-particle system within the strongly interacting regime, namely, where the pair is tightly bound. To this end, we expand up to second order in $\hat{H}_{0}$ the noninteracting two-particle Green's function $\hat{\mathcal{G}}_{0}(E) = (E-\hat{H}_{0})^{-1}$:
\begin{equation}
\hat{\mathcal{G}}_{0}(E) \simeq E^{-1} + E^{-1}\hat{H}_{0}E^{-1} + E^{-1}\hat{H}_{0}E^{-1}\hat{H}_{0}E^{-1},
\label{AEq1}
\end{equation}
the above expansion is accurate in the limit $|E|\gg J, \Delta$. As we will see shortly, this limit describes tightly bound dimers since $E\sim U$. To further proceed, it is convenient to write the two-body noninteracting Hamiltonian as $\hat{H}_{0} = \hat{T}\otimes\mathbb{I}_{1}+\mathbb{I}_{1}\otimes\hat{T}+\hat{V}\otimes\mathbb{I}_{1}+\mathbb{I}_{1}\otimes\hat{V}$, where $\hat{T}$ and $\hat{V}$ represent the hopping and the quasiperiodic potential, respectively:
\begin{equation}
\begin{split}
\hat{T} &= -J\sum_{i,j\neq i}\frac{1}{|i-j|^{\alpha}}|i\rangle\langle j|\\
\hat{V} &= \Delta\sum_{j}\cos(2\pi\beta j + \phi)|j\rangle\langle j|.
\end{split}
\label{AEq2}
\end{equation}
Evaluation of the matrix elements $\langle j, j|\hat{\mathcal{G}}_{0}(E)|i, i\rangle$ with $\hat{\mathcal{G}}_{0}(E)$ given in Eq. (\ref{AEq1}) gives:
\begin{equation}
\begin{split}
&\langle j, j|\hat{\mathcal{G}}_{0}(E)|i, i\rangle = \\
&\delta_{ij}\left[\frac{1}{E}+\frac{2V(i)}{E^{2}} + \frac{4J^{2}\zeta(2\alpha)+4V(i)^{2}}{E^3} \right]+\\
& \frac{2J^{2}}{E^{3}}\sum_{r}\frac{1}{r^{2\alpha}}[\delta_{n,m+r}+\delta_{n,m-r}]
\end{split}
\label{AEq3}
\end{equation}
being $\zeta(s)$ the Riemann zeta function. One can notice that if $\alpha\rightarrow\infty$ the previous equation reduces to the case of a quasicrystal with nearest neighbor hopping \cite{Orso1}. Substitution of Eq. (\ref{AEq3}) into Eq. (\ref{Eq5}) yields $E\sim U$ as a first approximation, which supports our previous assumptions. According to Eq. (\ref{AEq3}), the mobility of the bound pair emerges from second-order hopping processes, the amplitude of such tunneling follows a power-law $1/r^{\sigma}$ with $\sigma=2\alpha$ that is, twice the exponent of the individual particle hops.


\begin{thebibliography}{9}

\bibitem{Aubry1} S. Aubry and G. Andr\'e,
\href{}{Ann. Israel Phys. Soc. {\bf 3}, 133 (1980)}.

\bibitem{Macia1} E. Maci\'a,
\href{https://www.hindawi.com/journals/isrn/2014/165943/}{ISRN Condensed Matter Physics {\bf 2014}, 165943 (2014)}.

\bibitem{Sarma1} X. Li, X. Li, and S. Das Sarma,
\href{https://journals.aps.org/prb/abstract/10.1103/PhysRevB.96.085119}{Phys. Rev. B {\bf 96}, 085119 (2017)}.

\bibitem{Modugno1}
M. Modugno,
\href{https://iopscience.iop.org/article/10.1088/1367-2630/11/3/033023}{New J. Phys. {\bf 11}, 033023 (2009)}.

\bibitem{Tanatar1}
S. Roy, T. Mishra, B. Tanatar, and S. Basu,
\href{https://journals.aps.org/prl/abstract/10.1103/PhysRevLett.126.106803}{Phys. Rev. Lett. {\bf 126}, 106803 (2021)}.

\bibitem{Logan1}
A. Duthie, S. Roy, and D. E. Logan,
\href{https://journals.aps.org/prb/abstract/10.1103/PhysRevB.103.L060201}{Phys. Rev. B {\bf 103}, L060201 (2021)}.

\bibitem{Sarma2}
X. Li and S. Das Sarma,
\href{https://journals.aps.org/prb/abstract/10.1103/PhysRevB.101.064203}{Phys. Rev. B {\bf 101}, 064203 (2020)}.

\bibitem{Roati1} G. Roati, C. D'Errico, L. Fallani, M. Fattori, C. Fort, M. Zaccanti, G. Modugno, M. Modugno, and M. Inguscio,
\href{https://www.nature.com/articles/nature07071}{Nature {\bf 453}, 895-898 (2008)}.

\bibitem{Bloch1}
H. P. L\"uschen, S. Scherg, T. Kohlert, M. Schreiber, P. Bordia, X. Li, S. Das Sarma, and I. Bloch,
\href{https://journals.aps.org/prl/abstract/10.1103/PhysRevLett.120.160404}{Phys. Rev. Lett. {\bf 120}, 160404 (2018)}.

\bibitem{Lahini1}
Y. Lahini, R. Pugatch, F. Pozzi, M. Sorel, R. Morandotti, N. Davidson, and Y. Silberberg,
\href{https://journals.aps.org/prl/abstract/10.1103/PhysRevLett.103.013901}{Phys. Rev. Lett. {\bf 103}, 013901 (2009)}.

\bibitem{Harper1} P. G. Harper, 
\href{https://iopscience.iop.org/article/10.1088/0370-1298/68/10/304}{Proc. Phys. Soc. A {\bf 68}, 874 (1955)}.

\bibitem{Azbel1} M. Ya. Azbel,
\href{https://journals.aps.org/prl/abstract/10.1103/PhysRevLett.43.1954}{Phys. Rev. Lett. {\bf 43}, 1954 (1979)}.

\bibitem{Gustavo1}
G. A. Dominguez-Castro and R. Paredes,
\href{https://iopscience.iop.org/article/10.1088/1361-6404/ab1670}{Eur. J. Phys. {\bf 40}, 045403 (2019)}.

\bibitem{Macri1}
N. Defenu, T. Donner, T. Macr\`{\i}, G. Pagano, S. Ruffo, and A. Trombettoni,
\href{https://arxiv.org/abs/2109.01063}{arXiv:2109.01063 (2021)}.

\bibitem{DeMarco1} L. De Marco, G. Valtolina, K. Matsuda, W. G. Tobias, J. P. Covey, and J. Ye,
\href{http://dx.doi.org/10.1126/science.aau7230}{Science {\bf 363}, 853-856 (2019)}.

\bibitem{Moses1} S. A. Moses, J. P. Covey, M. T. Miecnikowski, D. S. Jin, and J. Ye
\href{https://www.nature.com/articles/nphys3985?proof=t}{Nature Physics {\bf 13}, 13-20 (2017)}.

\bibitem{Yan1} B. Yan, S. A. Moses, B. Gadway, J. P. Covey, K. R. A. Hazzard, A. M. Rey, D. S. Jin, and J. Ye, 
\href{https://www.nature.com/articles/nature12483}{Nature {\bf 501}, 521-525 (2013)}.

\bibitem{Browaeys1} A. Browaeys and T. Lahaye,
\href{https://www.nature.com/articles/s41567-019-0733-z}{Nature Physics {\bf 16}, 132-142 (2020)}.

\bibitem{Browaeys2} H. Labuhn, D. Barredo, S. Ravets, S. de L\'es\'eleuc, T. Macr\`{\i}, T. Lahaye, and A. Browaeys,
\href{https://www.nature.com/articles/nature18274}{Nature {\bf 534}, 667-670 (2016)}.

\bibitem{Blatt1} R. Blatt, C. F. Roos, 
\href{https://www.nature.com/articles/nphys2252}{Nature Physics {\bf 8}, 277-284 (2012)}.

\bibitem{Blatt2} P. Jurcevic, B. P. Lanyon, P. Hauke, C. Hempel, P. Zoller, R. Blatt, and C. F. Roos,
\href{https://www.nature.com/articles/nature13461}{Nature {\bf 511}, 202-205 (2014)}.

\bibitem{Monroe1} C. Monroe, W. C. Campbell, L.-M. Duan, Z.-X. Gong,  A. V. Gorshkov, P. W. Hess, R. Islam, K. Kim, N. M. Linke, G. Pagano, P. Richerme, C. Senko, and N. Y. Yao, 
\href{https://journals.aps.org/rmp/abstract/10.1103/RevModPhys.93.025001}{Rev. Mod. Phys. {\bf 93}, 025001 (2021)}.

\bibitem{Kimble1} C.-L. Hung, A. Gonz\'alez-Tudela, J. I. Cirac, and H. J. Kimble,
\href{https://www.pnas.org/doi/full/10.1073/pnas.1603777113}{Proc. Natl. Acad. Sci. USA {\bf 113}, E4946-E4955 (2016)}.

\bibitem{Santos1} X. Deng, S. Ray, S. Sinha, G. V. Shlyapnikov, and L. Santos, 
\href{https://journals.aps.org/prl/abstract/10.1103/PhysRevLett.123.025301}{Phys. Rev. Lett. {\bf 123}, 025301 (2019)}.

\bibitem{Roy1} N. Roy and A. Sharma, 
\href{https://journals.aps.org/prb/abstract/10.1103/PhysRevB.103.075124}{Phys. Rev. B {\bf 103}, 075124 (2021)}.

\bibitem{Santos2}  X. Deng, V. E. Kravtsov, G. V. Shlyapnikov, and L. Santos,
\href{https://journals.aps.org/prl/abstract/10.1103/PhysRevLett.120.110602}{Phys. Rev. Lett. {\bf 120}, 110602 (2018)}.

\bibitem{Scarola1} M. Yan, H.-Y. Hui, M. Rigol, and V. W. Scarola,
\href{https://journals.aps.org/prl/abstract/10.1103/PhysRevLett.119.073002}{Phys. Rev. Lett. {\bf 119}, 073002 (2017)}.

\bibitem{Panda1} R. K. Panda, A. Scardicchio, M. Schulz, S. R. Taylor, and M. {\v{Z}}nidari{\v{c}},
\href{https://iopscience.iop.org/article/10.1209/0295-5075/128/67003}{EPL (Europhysics Letters) {\bf 128}, 67003 (2019)}.

\bibitem{Vidmar1} J. {\v{S}}untajs, J. Bon{\v{c}}a, T. Prosen, and L. Vidmar,
\href{https://journals.aps.org/pre/abstract/10.1103/PhysRevE.102.062144}{Phys. Rev. E {\bf 102}, 062144 (2020)}.

\bibitem{Sirker1} M. Kiefer-Emmanouilidis, R. Unanyan, M. Fleischhauer, and J. Sirker,
\href{https://journals.aps.org/prl/abstract/10.1103/PhysRevLett.124.243601}{Phys. Rev. Lett. {\bf 124}, 243601 (2020)}.

\bibitem{Sirker2} M. Kiefer-Emmanouilidis, R. Unanyan, M. Fleischhauer, and J. Sirker,  
\href{https://journals.aps.org/prb/abstract/10.1103/PhysRevB.103.024203}{Phys. Rev. B {\bf 103}, 024203 (2021)}.

\bibitem{Huse1}
R. Nandkishore and D. A. Huse,
\href{https://www.annualreviews.org/doi/pdf/10.1146/annurev-conmatphys-031214-014726}{Annual Review of Condensed Matter Physics {\bf 6}, 15-38 (2015)}.

\bibitem{Vidal1}
G. Vidal,
\href{https://journals.aps.org/prl/abstract/10.1103/PhysRevLett.93.040502}{Phys. Rev. Lett. {\bf 93}, 040502 (2004)}.

\bibitem{Ulrich1}
S. Paeckel, T. K\"ohler, A. Swoboda, S. R. Manmana, U. Schollw\"ock, and C. Hubig,
\href{https://www.sciencedirect.com/science/article/pii/S0003491619302532?via%3Dihub}{Annals of Physics {\bf 411}, 167998 (2019)}.

\bibitem{Verstraete1}
J. Haegeman, J. I. Cirac, T. J. Osborne, I. Pi$\ifmmode \check{z}\else \v{z}\fi{}$orn , H. Verschelde, and F. Verstraete,
\href{https://journals.aps.org/prl/abstract/10.1103/PhysRevLett.107.070601}{Phys. Rev. Lett. {\bf 107}, 070601 (2011)}.

\bibitem{Bloch2} J.-y. Choi, S. Hild, J. Zeiher, P. Schau\ss{}, A. Rubio-Abadal, T. Yefsah, V. Khemani, D. A. Huse, I. Bloch, and C. Gross,
\href{https://www.science.org/doi/10.1126/science.aaf8834}{Science {\bf 352}, 1547-1552 (2016)}.

\bibitem{Bloch3} M. Schreiber, S. S. Hodgman, P. Bordia, H. P. L\"uschen, M. H. Fischer, R. Vosk, E. Altman, U. Schneider, and I. Bloch,
\href{https://www.science.org/doi/10.1126/science.aaa7432}{Science {\bf 349}, 842-845 (2015)}.

\bibitem{Bloch4} T. Kohlert, S. Scherg, X. Li, H. P. L\"uschen, S. Das Sarma, I. Bloch, and M. Aidelsburger,
\href{https://journals.aps.org/prl/abstract/10.1103/PhysRevLett.122.170403}{Phys. Rev. Lett. {\bf 122}, 170403 (2019)}.

\bibitem{Bloch5} A. Rubio-Abadal, J.-y. Choi, J. Zeiher, S. Hollerith, J. Rui, I. Bloch, and C. Gross,
\href{https://journals.aps.org/prx/abstract/10.1103/PhysRevX.9.041014}{Phys. Rev. X {\bf 9}, 041014 (2019)}.

\bibitem{Shepelyansky1} D. L. Shepelyansky,
\href{https://journals.aps.org/prl/abstract/10.1103/PhysRevLett.73.2607}{Phys. Rev. Lett. {\bf 73}, 2607 (1994)}.

\bibitem{Oppen1} F. von Oppen, T. Wettig, and J. M\"uller,
\href{https://journals.aps.org/prl/abstract/10.1103/PhysRevLett.76.491}{Phys. Rev. Lett. {\bf 76}, 491 (1996)}.

\bibitem{Flach1} D. O. Krimer, R. Khomeriki, and S. Flach,
\href{https://link.springer.com/article/10.1134/S0021364011170097}{JETP Letters {\bf 94}, 406-412 (2011)}.

\bibitem{Flach2} D. Thongjaomayum, A. Andreanov, T. Engl, and S. Flach,
\href{https://journals.aps.org/prb/abstract/10.1103/PhysRevB.100.224203}{Phys. Rev. B {\bf 100}, 224203 (2019)}.

\bibitem{Orso1} G. Dufour and G. Orso,
\href{https://journals.aps.org/prl/abstract/10.1103/PhysRevLett.109.155306}{Phys. Rev. Lett. {\bf 109}, 155306 (2012)}.

\bibitem{Flach3} S. Flach, M. Ivanchenko, and R. Khomeriki,
\href{https://iopscience.iop.org/article/10.1209/0295-5075/98/66002}{EPL {\bf 98}, 66002 (2012)}.

\bibitem{Frahm1} K. M. Frahm and D. L. Shepelyansky,
\href{https://link.springer.com/article/10.1140/epjb/e2015-60733-9}{Eur. Phys. J. B {\bf 88}, 337 (2015)}.

\bibitem{Shepelyansky2} D. L. Shepelyansky,
\href{https://journals.aps.org/prb/abstract/10.1103/PhysRevB.54.14896}{Phys. Rev. B {\bf 54}, 14896 (1996)}.

\bibitem{Kim1} P. H. Song and D. Kim,
\href{https://journals.aps.org/prb/abstract/10.1103/PhysRevB.56.12217}{Phys. Rev. B {\bf 56}, 12217 (1997)}.

\bibitem{Mujal1} P. Mujal, A. Polls, S. Pilati, and B. Juli\'a-D\'iaz,
\href{https://journals.aps.org/pra/abstract/10.1103/PhysRevA.100.013603}{Phys. Rev. A {\bf 100}, 013603 (2019)}.

\bibitem{Liu1} G. Liu,
\href{https://link.springer.com/article/10.1140/epjb/s10051-020-00036-0}{Eur. Phys. J. B. {\bf 94}, 12 (2021)}.

\bibitem{Rai1} C. Lee, A. Rai, C. Noh, and D. G. Angelakis,
\href{https://journals.aps.org/pra/abstract/10.1103/PhysRevA.89.023823}{Phys. Rev. A {\bf 89}, 023823 (2014)}.

\bibitem{Orso2} F. Stellin and G. Orso,
\href{https://journals.aps.org/prresearch/abstract/10.1103/PhysRevResearch.2.033501}{Phys. Rev. Research {\bf 2}, 033501 (2020)}.

\bibitem{Orso3} F. Stellin and G. Orso,
\href{https://journals.aps.org/prb/abstract/10.1103/PhysRevB.102.144201}{Phys. Rev. B {\bf 102}, 144201 (2020)}.

\bibitem{Andreanov1} D. Thongjaomayum, S. Flach, and A. Andreanov,
\href{https://journals.aps.org/prb/abstract/10.1103/PhysRevB.101.174201}{Phys. Rev. B {\bf 101}, 174201 (2020)}.

\bibitem{Gustavo2} G. A. Dom\'inguez-Castro and R. Paredes,
\href{https://journals.aps.org/pra/abstract/10.1103/PhysRevA.104.033306}{Phys. Rev. A {\bf 104}, 033306 (2021)}.

\bibitem{Jochim1} F. Serwane, G. Z\"urn, T. Lompe, T. B. Ottenstein, A. N. Wenz, and S. Jochim,
\href{https://www.science.org/doi/10.1126/science.1201351}{Science {\bf 332}, 336-338 (2011)}.

\bibitem{Jochim2} A. Bergschneider, V. M. Klinkhamer, J. H. Becher, R. Klemt, L. Palm, G. Z\"urn, S. Jochim, and P. M. Preiss,
\href{https://www.nature.com/articles/s41567-019-0508-6}{Nature Physics {\bf 15}, 640-644 (2019)}.

\bibitem{Jochim3} J. H. Becher, E. Sindici, R. Klemt, S. Jochim, A. J. Daley, and P. M. Preiss,
\href{https://journals.aps.org/prl/abstract/10.1103/PhysRevLett.125.180402}{Phys. Rev. Lett. {\bf 125}, 180402 (2020)}.

\bibitem{Macri2} 
T. Macr\`{\i}, L. Lepori, G. Pagano, M. Lewenstein, and L. Barbiero,
\href{https://journals.aps.org/prb/abstract/10.1103/PhysRevB.104.214309}{Phys. Rev. B {\bf 104}, 214309 (2021)}.

\bibitem{Santos3}  W. Li, A. Dhar, X. Deng, K. Kasamatsu, L. Barbiero, and L. Santos,
\href{https://journals.aps.org/prl/abstract/10.1103/PhysRevLett.124.010404}{Phys. Rev. Lett. {\bf 124}, 010404 (2020)}.

\bibitem{Santos4}  W.-H. Li, A. Dhar, X. Deng, and L. Santos,
\href{https://journals.aps.org/pra/abstract/10.1103/PhysRevA.103.043331}{Phys. Rev. A {\bf 103}, 043331 (2021)}.

\bibitem{Wilkinson1}  M. Wilkinson,
\href{https://royalsocietypublishing.org/doi/10.1098/rspa.1984.0016}{Proc. R. Soc. Lond. A {\bf 391}, 305-350 (1984)}.

\bibitem{Martin1} A. T. Sommer, L. W. Cheuk, M. J. H. Ku, W. S. Bakr, and M. Zwierlein,
\href{https://journals.aps.org/prl/abstract/10.1103/PhysRevLett.108.045302}{Phys. Rev. Lett. {\bf 108}, 045302 (2012)}.

\bibitem{Tilman1} H. Moritz, T. St\"oferle, K. G\"unter, M. K\"ohl, and T. Esslinger,
\href{https://journals.aps.org/prl/abstract/10.1103/PhysRevLett.94.210401}{Phys. Rev. Lett. {\bf 94}, 210401 (2005)}.

\bibitem{Saha1} M. Saha, S. K. Maiti, and A. Purkayastha,
\href{https://journals.aps.org/prb/abstract/10.1103/PhysRevB.100.174201}{Phys. Rev. B {\bf 100}, 174201 (2019)}.
\end{thebibliography}
\end{document}